\documentclass[a4paper, amsfonts, amssymb, amsmath, reprint, showkeys, nofootinbib, twoside,notitlepage,onecolumn]{revtex4-2}

\def\prd{{Phys.~Rev.~D}}        
\def\prl{{Phys.~Rev.~Lett.}}    
\usepackage{amsmath,amstext}
\usepackage[T1]{fontenc}
\usepackage{amssymb}
\input{epsf}
\usepackage{graphicx}
\usepackage{ae,aecompl}

\usepackage{hyperref}
\usepackage{amsmath}
\usepackage{amssymb}
\usepackage{mathtools}
\usepackage{bm}
\usepackage{cleveref}
\usepackage{tensor}
\usepackage{braket}
\usepackage{enumitem}
\usepackage{mhchem}
\usepackage{cancel}
\usepackage{amsthm}
\usepackage{upgreek}
\usepackage{mathrsfs}

\usepackage{color}

\newcommand{\beqy}{\begin{eqnarray}}
\newcommand{\eeqy}{\end{eqnarray}}

\def\be{\begin{equation}}
\def\ee{\end{equation}}
\def\beq{\begin{eqnarray}}
\def\eeq{\end{eqnarray}}

\theoremstyle{definition}

\theoremstyle{theorem}

\theoremstyle{corollary}

\newcommand{\mn}{{\mu\nu}}

\crefname{equation}{}{} 

\begin{document}
\title{Relativistic superfluid profiles near critical surfaces}
\author{L.~Gavassino$^1$ and A.~Soloviev$^2$}
\affiliation{$^1$Department of Applied Mathematics and Theoretical Physics, University of Cambridge, Wilberforce Road, Cambridge CB3 0WA, United Kingdom\\
$^2$Faculty of Mathematics and Physics, University of Ljubljana, Jadranska ulica 19, SI-1000, Ljubljana, Slovenia}

\begin{abstract}
Landau’s two-fluid model of superfluidity ceases to apply in regions where the condensate amplitude exhibits rapid spatial variation, such as vortex cores or in the vicinity of container walls. A recently proposed relativistic Gross--Pitaevskii--type framework treats the condensate as an independent scalar degree of freedom, enabling a controlled analysis of such regimes. We use it to construct stationary superflows close to the superfluid--normal phase boundary, and examine their stability. We obtain an exact expression for Landau’s critical velocity and show that the standard Newtonian profiles (such as the near-vortex condensate depletion or the boundary-layer decay) persist unmodified in the relativistic setting. We further analyse a genuinely relativistic configuration in which an accelerated superfluid develops a phase boundary induced by Tolman temperature gradients.
\end{abstract} 

\maketitle

\section{Introduction}

A superfluid is a fluid substance that, due to 
spontaneous symmetry breaking, possesses a macroscopic order parameter, in the form of a complex scalar field $\Sigma=\sigma e^{i\psi}$ (with $\sigma,\psi\in \mathbb{R}$). For a superfluid in local equilibrium, the phase $\psi$ undergoes rapid oscillations, on a timescale set by $\mu^{-1}$ (with $\mu$ the chemical potential). Owing to these fast dynamics, Landau proposed that a non-dissipative hydrodynamic description of a superfluid should be formulated in terms of the two vector fields $\beta^\mu$ and $\nabla^\mu\psi$ \cite[\S 139]{landau6}. Here, $\beta^\mu$ denotes the inverse-temperature four-vector \cite{VanKampen1968,Israel_Stewart_1979,cool1995,BecattiniBeta2016,GavassinoTermometri,Israel_2009_inbook} that specifies the local equilibrium ensemble of the excitations,\footnote{In any non-dissipative hydrodynamic theory, the system is assumed to be in local equilibrium, meaning that the probability distribution for the excitations takes the canonical form $e^{\beta^\mu p_\mu}$, with $p_\mu$ the four-momentum of the state.} while $\nabla^\mu\psi$ captures the dynamics of the condensate's phase. The resulting hydrodynamic theory is known as ``Landau's two-fluid model'' (or Tisza's two-fluid model \cite{huang_book}), since it describes a substance that supports two distinct types of macroscopic motion, both of which take place simultaneously and at the same location. The associated four velocities are
\begin{equation}\label{normalandsuperf}
u^\mu = \frac{\beta^\mu}{\sqrt{-\beta^\nu \beta_\nu}}\,, \qquad \qquad 
w^\mu = \frac{\nabla^\mu\psi}{\sqrt{-\nabla^\nu\psi \nabla_\nu\psi}}\,,
\end{equation}
where $u^\mu$ is the \emph{normal} velocity (shared with all conventional fluids), whereas $w^\mu$ defines the \emph{superfluid} velocity, associated with a new (frictionless) component of the flow \cite{khalatnikov_book,Termo}.

The two-fluid framework provides an efficient macroscopic description of superfluidity \cite{Carter_starting_point,Gusakov:2016eom,GavassinoIordanskii2021}, yet it becomes inadequate in certain regimes. Its central limitation lies in the assumption that the amplitude $\sigma$ of the order parameter is fixed algebraically by the local thermodynamic variables, namely $\sigma=\bar{\sigma}(\beta\!\cdot\!\beta,\beta\!\cdot\nabla\psi,\nabla\psi\!\cdot\nabla\psi)$. In a homogeneous equilibrium configuration, the expectation value $\bar{\sigma}$ is uniquely defined, corresponding to the condensate sitting at the minimum of a symmetry-breaking potential, for instance of the Mexican-hat form $V\sim(\sigma^{2}-\bar{\sigma}^{2})^{2}$. In inhomogeneous equilibria, however, the free energy contains an additional gradient term, $\nabla_\mu\sigma\,\nabla^\mu\sigma$, which competes with the potential and drives $\sigma$ away from $\bar{\sigma}$. Consequently, Landau’s two-fluid description ceases to be reliable once spatial gradients of $\sigma$ become appreciable. This occurs, for instance, in the vicinity of a vortex core \cite[\S 30]{landau9} or in regions where the system undergoes a rapid superfluid–normal transition \cite[\S 55]{FetterWalecka1971}, like near the chiral phase transition \cite{Grossi:2020ezz,Grossi:2021gqi}. A consistent treatment of such configurations requires a ``Gross–Pitaevskii–type'' approach, in which the full complex order parameter $\Sigma$ (rather than solely its phase) is promoted to a dynamical degree of freedom. A relativistic finite-temperature formulation of this kind has, in fact, been developed quite recently \cite{Mitra:2020hbj,Buza:2024jxe}, building on the Son-Nicolis formalism \cite{Son:2002zn,Nicolis:2011cs}.
  
In this work, we employ the relativistic Gross–Pitaevskii theory of \cite{Mitra:2020hbj,Buza:2024jxe} to characterize the equilibrium structure of relativistic superfluids in regimes where $\sigma$ exhibits significant spatial variation. In addition to analyzing the single-vortex configuration and the boundary layer, we examine equilibria in which strong accelerations generate a Tolman–induced temperature gradient, producing a stratified region across which the system transitions rapidly from the superfluid to the normal phase. 

Throughout the article, we adopt the metric signature $(-,+,+,+)$, and work in natural units, $c\,{=}\,\hbar\,{=}\,k_B\,{=}\,1$. The spacetime, albeit curved in general, is treated as a fixed background. All indices introduced in the text ($\mu$, $I$, $h$...) obey Einstein's convention, and square brackets around indices denote antisymmetrization: $A_{[\mn]}=\frac{1}{2}\left(A_{\mn}-A_{\nu\mu}\right)$.

\section{Outline of the hydrodynamic model}
\vspace{-0.2cm}

In this section, we provide a brief overview of the hydrodynamic theory of \cite{Mitra:2020hbj,Buza:2024jxe}, in the non-dissipative limit. The theory can be derived from an action principle. Here, we sidestep the derivation, and merely list the final equations.

\vspace{-0.2cm}
\subsection{Constitutive relations}
\vspace{-0.2cm}

The dynamical fields of the theory are $\beta^\mu$, $\mu$, and the complex order parameter $\Sigma=\sigma e^{i\psi}$. From the first, one obtains the temperature, $T=(-\beta_\mu\beta^\mu)^{-1/2}$, and the normal velocity $u^\mu$ via \eqref{normalandsuperf}. In contrast to Landau’s two–fluid framework, $\mu$ and $\Sigma$ here are treated as \textit{independent variables}, and the Josephson relation $\mu=-u^\mu\nabla_\mu\psi$ is \emph{not} imposed at the outset. The thermodynamic input of the model is encoded in the expression for the total pressure,
\begin{equation}
P=p(T,\mu,\sigma)-\dfrac{1}{2}\nabla_\mu \sigma \nabla^\mu \sigma -\dfrac{1}{2} \sigma^2 \nabla_\mu \psi \nabla^\mu \psi \, ,
\end{equation}
where $p$ denotes a generic function of its arguments. The differential $dp=sdT+nd\mu+Fd\sigma$ identifies the entropy density $s$, the normal density $n$, and the force $F$ conjugate to variations of $\sigma$.

Given the above notation, the constitutive relations for the stress-energy tensor $T^{\mu \nu}$, the particle current $J^\mu$, and the entropy current $s^\mu$, are given by
\begin{equation}\label{constitutiverelation}
\begin{split}
T^{\mu \nu}={}& P g^{\mu \nu}+(Ts{+}\mu n)u^\mu u^\nu +\nabla^\mu \sigma \nabla^\nu \sigma +\sigma^2 \nabla^\mu \psi \nabla^\nu \psi \, , \\
J^{\mu}={}& nu^\mu +\sigma^2 \nabla^\mu \psi \, , \\
s^{\mu}={}& su^\mu \, . \\
\end{split}
\end{equation}

\vspace{-0.7cm}
\subsection{Equations of motion}
\vspace{-0.2cm}

The degrees of freedom $\{\beta^\mu,\mu,\Sigma\}$ constitute a total of 7 independent (real) field components. Hence, we need 7 equations of motion. Writing down the conservation laws explicitly $\nabla_\mu T^{\mu \nu}=\nabla_\mu J^{\mu}= 0$, one finds that the only way to avoid overdetermination and enforce reversibility is to impose the following 7 independent equations:
\begin{equation}\label{EoMs}
\begin{split}
& (Ts{+}\mu n)u^\mu \nabla_\mu u^\nu +(g^{\nu \mu}{+}u^\nu u^\mu)(s\nabla_\mu T +n\nabla_\mu \mu) =0 \, , \qquad \nabla_\mu (su^\mu)=0 \, , \qquad \nabla_\mu (nu^\mu)=0 \, , \\
&\nabla_\mu (\sigma^2 \nabla^\mu \psi) =0 \, , \\
& \nabla_\mu \nabla^\mu \sigma +F -\sigma \nabla_\mu \psi \nabla^\mu \psi =0 \, . \\
\end{split}
\end{equation}
The first line is the Euler equations for the normal component. The second line enforces conservation for the superfluid particle current (in Landau's two-fluid model, only the \textit{total} current $J^\mu$ is conserved). The third line is the evolution equation for $\sigma$. This completes the construction. In Appendix \ref{causalone}, we compute the propagation speeds of the model.

\vspace{-0.3cm}
\subsection{Condensate amplitude in uniform equilibria}\label{Unofrmia}
\vspace{-0.3cm}

\begin{figure}
    \centering
\includegraphics[width=0.47\linewidth]{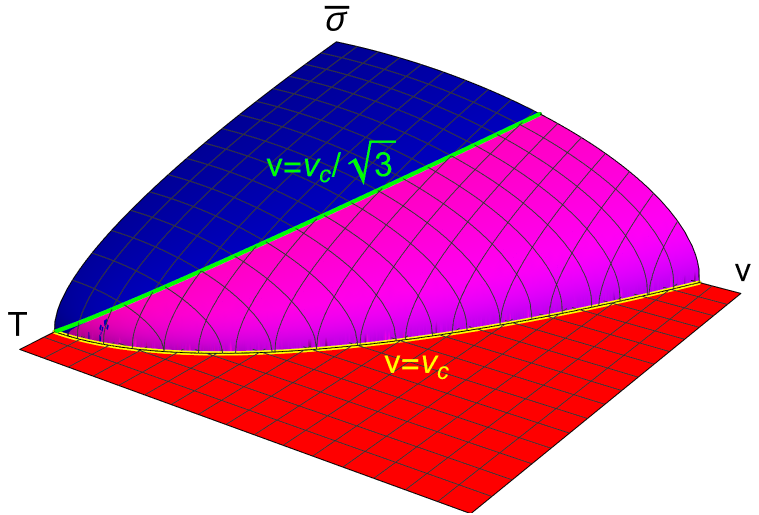}
\caption{Phase diagram of the equilibrium condensate amplitude $\bar{\sigma}(T,v)$ at fixed $\mu$ for a fluid governed by the equation of state \eqref{EoS}. The superfluid region (where $\bar{\sigma}\neq 0$) is shown in blue, while the normal phase is indicated in red. We will show in section \ref{LandauIsBack} that there is an intermediate magenta band $v_c/\sqrt{3}\, {\le}\, v\, {\le}\, v_c$, with $v_c\,{=}\,\sqrt{m_0(T_c{-}T)/\mu^2}$, in which the naive maximization of $P(\sigma)$ predicts a nonzero condensate, yet the corresponding configurations are unstable to fluctuations. This instability implies that superfluidity actually breaks down at $v_c/\sqrt{3}$, which therefore represents Landau's critical velocity in this model.}
    \label{fig:Uniform}
\end{figure}

We recall that our primary objective is to determine how spatial inhomogeneities drive the condensate amplitude $\sigma$ away from the expectation value $\bar{\sigma}$ characteristic of uniform equilibrium configurations. It is therefore useful to obtain an explicit expression for $\bar{\sigma}$. Following \cite{Buza:2024jxe}, we assume that the function, $p,$ has a Mexican-hat form\footnote{Note that, in this article, we have replaced the gauge covariant derivative $D_\mu\psi$ \cite{Mitra:2020hbj,Buza:2024jxe} with the usual covariant derivative $\nabla_\mu \psi$. For this reason, we needed to add a $\mu^2$ correction to the quadratic term in $\sigma$, which would otherwise be contained in the kinetic term.}:
\begin{equation}\label{EoS}
p(T,\mu,\sigma)=p(T,\mu,0)+\dfrac{m_0(T_c{-}T){-}\mu^2}{2}\sigma^2 -\dfrac{\lambda}{4}\sigma^4 \, ,
\end{equation}
where $T_c$ (the critical temperature), $m_0$, and $\lambda$ are strictly positive constants\footnote{Note that we are inspired by e.g.~spontaneous magnetism, which grows below the Curie temperature like $m\sim (T-T_c)^{1/2}$. Different scaling exponents can be considered, depending on the physical system of interest.}. In a homogeneous equilibrium state we have $\sigma=\mathrm{const}\equiv\bar{\sigma}$, and the fourth line of \eqref{EoMs} reduces to the stationary-point condition  
$0=F-\sigma\nabla_\mu \psi \nabla^\mu \psi \equiv \partial_\sigma P$.
Solving for $\sigma$, and taking the solution that maximizes $P$ (in equilibrium, the grand potential $\Omega=-PV$ is minimized), we obtain $\Bar{\sigma}$. Introducing $v^2=1-(u^\mu w_\mu)^{-2}$ as the relative speed between the normal and superfluid components, and considering that the Josephson relation $u_\mu \nabla^\mu \psi=-\mu$ should hold in equilibrium, one finds
\begin{equation}\label{nereee}
\bar{\sigma}(T,\mu,v)=
\begin{cases}
\sqrt{\dfrac{m_0 (T_c-T)-\mu^2v^2}{\lambda}} & \text{ if } m_0 T_c\geq m_0 T+\mu^2v^2 \, , \\
0 & \text{ otherwise} \, .\\
\end{cases}
\end{equation}
This equation tells us that the condensate is destroyed either if the superfluid is above the critical temperature, or if $v$ exceeds some maximal velocity (see figure \ref{fig:Uniform}), in agreement with the standard theory of superfluidity.

\section{Equilibrium conditions}
\vspace{-0.2cm}

In this section, we follow the same procedure as in \cite{GavassinoGENERIC2022,Gavassino:2023qnw,GavassinoSuperflowsStable2025ldj} to identify the class of (possibly non-uniform) configurations that effectively behave as equilibrium solutions of the theory. By an ``equilibrium solution’’ we mean a field configuration satisfying the equations of motion \eqref{EoMs} and the following conditions:
\begin{itemize}
\item[\textbf{(a)}] It is stationary, i.e.\ invariant under a group of spacetime isometries whose orbits are timelike.
\item[\textbf{(b)}] It is stable against linear perturbations.
\item[\textbf{(c)}] It serves as a local late-time attractor when the system is weakly coupled to a heat bath.
\end{itemize}
As we shall demonstrate shortly, a superfluid exhibits the remarkable feature that, for a fixed heat bath, several distinct configurations can satisfy conditions \textbf{(a,b,c)} simultaneously. These represent states in which the normal component comoves with the bath, while the superfluid component executes an independent stationary flow. In the thermodynamic limit, such configurations acquire arbitrarily long lifetimes.

\subsection{Thermodynamically admissible processes}

Consider coupling the superfluid weakly to a heat and particle bath, so that the combined ``superfluid + bath'' system is isolated. Any admissible process must obey the second law, $\Delta S {+} \Delta S^{B} {\ge} 0$, and must conserve all Noether charges of the microscopic theory (energy–momentum, particle number, etc.), $\Delta Q_I {+} \Delta Q_I^{B} {=} 0$. Here and below, the superscript $B$ denotes bath quantities, and $I$ is a charge index. Assuming that the bath is always in internal equilibrium and is much larger than the superfluid, its entropy variation is $\Delta S^{B} = - \alpha_\star^{I}\, \Delta Q_I^{B} = \alpha_\star^{I}\, \Delta Q_I$, where the intensive variables $\alpha_\star^{I}$ are effectively constant. Consequently, the condition for thermodynamic admissibility becomes $\Delta\!\left(S + \alpha_\star^{I} Q_I\right) \ge 0$.

The argument above implies that, as the combined system ``superfluid + bath'' evolves, the functional
\begin{equation}
\Phi[\beta^\nu,\mu,\Sigma] = S[\beta^\nu,\mu,\Sigma] + \alpha_\star^{I} Q_I[\beta^\nu,\mu,\Sigma]
\end{equation}
(with $\alpha_\star^{I}$ treated as fixed external parameters) increases monotonically until it reaches a local maximum. Once there, any further evolution would necessarily reduce $\Phi$, and is therefore prohibited by thermodynamics. Consequently, the local maxima of $\Phi$ are stationary configurations (as they do not evolve), are  Lyapunov stable under small perturbations (since any departure would lower $\Phi$) \cite{lasalle1961stability}, and act as local attractors (because nearby configurations relax toward them) \cite{GavassinoGibbs2021}. They thus satisfy all criteria \textbf{(a,b,c)}, and can be regarded as equilibrium states\footnote{Full thermodynamic equilibrium corresponds to the \textit{global} maximum of $\Phi$, and typically describes a state in which the superfluid component comoves with the normal component \cite{GavassinoStabilityCarter2022} (when this is consistent with the requirement that $\nabla\psi$ be a closed one-form). This configuration is the unique global attractor once one allows for rare, macroscopically large thermal fluctuations that can momentarily violate the second law. By contrast, the local maxima of $\Phi$ discussed here should be understood as metastable states \cite[13.5]{huang_book}, whose lifetime diverges in the thermodynamic limit, where macroscopic violations of the second law become infinitely unlikely. For an in-depth discussion, see~\cite{GavassinoSuperflowsStable2025ldj}.

}.

\vspace{-0.1cm}
\subsection{Explicit formula for the thermodynamic potential}
\vspace{-0.2cm}

The functional $\Phi=S+\alpha_\star^I Q_I$ can be expressed explicitly as
\begin{equation}\label{PhiasaFlux}
\Phi=\int_{\text{``Cauchy surface''}} \phi^\mu\, d^3{\mathbf{\Sigma}}_\mu \, ,
\end{equation}
where $\phi^\mu = s^\mu + \alpha_\star^{I} J_I^\mu$. Here, $s^\mu$ and $J_I^\mu$ denote, respectively, the entropy current and the Noether currents of the superfluid, while we recall that $\alpha_\star^{I}=\text{const}$ are fixed parameters characterizing the external bath. For the present system, the relevant Noether currents consist of the particle current $J^\mu$ together with the currents $\mathcal{K}_h^{\nu} T^\mu_{\ \nu}$ associated with the spacetime Killing vectors $\mathcal{K}_h^{\nu}$ (recall that the spacetime geometry is regarded as nondynamical) \cite[3.2]{HawkingEllis1973}. Accordingly,
$\phi^\mu \,{=}\, s^\mu \,{+}\, \alpha_\star J^\mu \,{+}\, \alpha_\star^{h}\, \mathcal{K}_h^{\nu} T^\mu_{\ \nu}$, with $s^\mu$, $J^\mu$, and $T^{\mu}_\nu$ given by \eqref{constitutiverelation}. Introducing the composite Killing vector $\beta_\star^{\nu} {\equiv} \alpha_\star^{h} \mathcal{K}_h^{\nu}$, this expression takes the compact form
\begin{equation}
\phi^\mu = s^\mu + \alpha_\star J^\mu + \beta_\star^{\nu} T^\mu_{\ \nu} \, .
\end{equation}

\vspace{-0.2cm}
\subsection{Stationary points}
\vspace{-0.2cm}

To locate the maxima of $\Phi$, we consider a one-parameter family of solutions $\{\beta^\nu(\epsilon),\mu(\epsilon),\Sigma(\epsilon)\}_{\epsilon\in\mathbb{R}}$ to the equations of motion \eqref{EoMs}, with $\epsilon=0$ representing the candidate equilibrium state. We then impose the conditions $\dot{\Phi}(0)=0$ and $\ddot{\Phi}(0)\le 0$ (with $\dot{f}\equiv df/d\epsilon$). The first requirement identifies the flow configurations for which $\Phi$ is stationary, while the second yields the thermodynamic inequalities necessary for the configuration to constitute a true local maximum.

Using the constitutive relations \eqref{constitutiverelation}, we find\footnote{Since the spacetime geometry and the bath parameters are regarded as fixed external data, the quantities $g^{\mu\nu}$, $\alpha_\star$, and $\beta_\star^\nu$ must be held constant when taking derivatives with respect to $\epsilon$. The subscript ``$\star$’’ serves as a reminder of this fact.}
\begin{equation}\label{theBigMess}
\begin{split}
\phi^{\mu}={}& \beta_\star^\nu \nabla_\nu \sigma \nabla^\mu \sigma +\left[p-\tfrac{1}{2}\nabla_\nu \sigma\nabla^\nu \sigma-\tfrac{1}{2} \sigma^2 \nabla_\nu \psi \nabla^\nu \psi \right]\beta_\star^\mu +\left[s{+}\alpha_\star n +u_\nu \beta_\star^\nu (Ts{+}\mu n)\right]u^\mu+\left[\alpha_\star {+}\beta_\star^\nu \nabla_\nu \psi\right]\sigma^2 \nabla^\mu \psi \, , \\
\Dot{\phi}^{\mu}={}& \beta_\star^\nu \nabla_\nu \Dot{\sigma} \nabla^\mu \sigma+\beta_\star^\nu \nabla_\nu \sigma \nabla^\mu \Dot{\sigma}+ \left[s\Dot{T}+n\Dot{\mu}+F\Dot{\sigma}-\nabla_\nu \sigma \nabla^\nu \Dot{\sigma}- \sigma \Dot{\sigma} \nabla_\nu \psi \nabla^\nu \psi-\sigma^2 \nabla_\nu \psi \nabla^\nu \Dot{\psi} \right]\beta_\star^\mu  \\
+{}&\left[s{+}\alpha_\star n +u_\nu \beta_\star^\nu (Ts{+}\mu n)\right]\Dot{u}^\mu +\left[\Dot{s}{+}\alpha_\star \Dot{n} +\Dot{u}_\nu \beta_\star^\nu (Ts{+}\mu n)+u_\nu \beta_\star^\nu (\Dot{T}s{+}T\Dot{s}{+}\Dot{\mu} n{+}\mu \Dot{n})\right]u^\mu  \\
+{}&\beta_\star^\nu \nabla_\nu \Dot{\psi}\,\sigma^2 \nabla^\mu \psi+\left[\alpha_\star {+}\beta_\star^\nu \nabla_\nu \psi\right]\left[2\sigma\Dot{\sigma} \nabla^\mu \psi+\sigma^2 \nabla^\mu \Dot{\psi}\right],\\
\Ddot{\phi}^{\mu}={}& \beta_\star^\nu \nabla_\nu \Ddot{\sigma} \nabla^\mu \sigma+2\beta_\star^\nu \nabla_\nu \Dot{\sigma} \nabla^\mu \Dot{\sigma}+\beta_\star^\nu \nabla_\nu \sigma \nabla^\mu \Ddot{\sigma}+ \left[\Dot{s}\Dot{T}+s\Ddot{T}+\Dot{n}\Dot{\mu}+n\Ddot{\mu}+\Dot{F}\Dot{\sigma}+F\Ddot{\sigma}\right] \beta_\star^\mu \\ 
-{}&\left[\nabla_\nu \Dot{\sigma} \nabla^\nu \Dot{\sigma}+\nabla_\nu \sigma \nabla^\nu \Ddot{\sigma}+ \Dot{\sigma}^2 \nabla_\nu \psi \nabla^\nu \psi+ \sigma \Ddot{\sigma} \nabla_\nu \psi \nabla^\nu \psi+ 4\sigma \Dot{\sigma} \nabla_\nu \psi \nabla^\nu \Dot{\psi}+\sigma^2 \nabla_\nu \Dot{\psi} \nabla^\nu \Dot{\psi}+\sigma^2 \nabla_\nu \psi \nabla^\nu \Ddot{\psi} \right]\beta_\star^\mu\\
+{}&\left[s{+}\alpha_\star n +u_\nu \beta_\star^\nu (Ts{+}\mu n)\right]\Ddot{u}^\mu+2\left[\Dot{s}{+}\alpha_\star \Dot{n} +\Dot{u}_\nu \beta_\star^\nu (Ts{+}\mu n)+u_\nu \beta_\star^\nu (\Dot{T}s{+}T\Dot{s}{+}\Dot{\mu} n{+}\mu \Dot{n})\right]\Dot{u}^\mu  \\
+{}&\left[\Ddot{s}{+}\alpha_\star \Ddot{n} +\Ddot{u}_\nu \beta_\star^\nu (Ts{+}\mu n)+2\Dot{u}_\nu \beta_\star^\nu (\Dot{T}s{+}T\Dot{s}{+}\Dot{\mu} n{+}\mu \Dot{n})+u_\nu \beta_\star^\nu (\Ddot{T}s{+}2\Dot{T}\Dot{s}{+}T\Ddot{s}{+}\Ddot{\mu} n{+}2\Dot{\mu}\Dot{n}{+}\mu \Ddot{n})\right]u^\mu\\
+{}&\beta_\star^\nu \nabla_\nu \Ddot{\psi}\,\sigma^2 \nabla^\mu \psi+2\beta_\star^\nu \nabla_\nu \Dot{\psi}\,\left[2\sigma\Dot{\sigma} \nabla^\mu \psi{+}\sigma^2 \nabla^\mu \Dot{\psi}\right]+\left[\alpha_\star {+}\beta_\star^\nu \nabla_\nu \psi\right]\left[2\Dot{\sigma}^2 \nabla^\mu \psi{+}2\sigma\Ddot{\sigma} \nabla^\mu \psi{+}4\sigma\Dot{\sigma} \nabla^\mu \Dot{\psi}{+}\sigma^2 \nabla^\mu \Ddot{\psi}\right].\\
\end{split}
\end{equation}
It is straightforward to verify that, if we require $\dot{\Phi}(0)=\int \dot{\phi}^\mu(0)\, d^3\mathbf{\Sigma}_\mu$ to vanish for arbitrary Cauchy surfaces and for an arbitrary choice of the family $\{\beta^\nu(\epsilon),\mu(\epsilon),\Sigma(\epsilon)\}_{\epsilon\in\mathbb{R}}$, then the terms $\beta_\star^\nu \nabla_\nu \sigma$, $s+\alpha_\star n+u_\nu \beta_\star^\nu (Ts+\mu n)$, $\Dot{u}_\nu \beta_\star^\nu$, and $\alpha_\star+\beta_\star^\nu \nabla_\nu \psi$ must all vanish individually. Imposing this condition, together with requirement \textbf{(a)}, yields the following equilibrium relations on the ``$\epsilon=0$'' state\footnote{Note that the $U(1)$ symmetry $\psi \rightarrow \psi+\text{``constant''}$ implies that while $\nabla\psi$ is observable, the local value of the phase $\psi$ itself is not. Hence, while we must set $\mathfrak{L}_{\beta_\star}\nabla \psi = 0$, the derivative $\mathfrak{L}_{\beta_\star}\psi=\beta_\star^\nu \nabla_\nu \psi=-\alpha_\star$ is in general non-zero.}:
\begin{equation}\label{equilibriamo}
\frac{u^\mu}{T}=\beta_\star^\mu \, , 
\qquad 
\frac{\mu}{T}=\alpha_\star=-\beta_\star^\nu \nabla_\nu \psi \, , 
\qquad 
\mathfrak{L}_{\beta_\star}\text{``Observables''}=0 \, .
\end{equation}
We see that the Josephson relation $u^\mu \nabla_\mu \psi=-\mu$ is recovered as an equilibrium condition, as expected.
The other conditions are standard: $u^\mu/T$ is a Killing vector \cite{BecattiniBeta2016} and $\mu/T$ is constant, both matching the bath values, as required by the zeroth law of thermodynamics \cite{GavassinoTermometri}.

Conditions \eqref{equilibriamo} are also fully consistent with the first line of \eqref{EoMs}. In fact, the Killing condition immediately implies that $T u^\mu\nabla_\mu u^\nu + (g^{\nu\mu}{+}u^\nu u^\mu)\nabla_\mu T = 0$. Since $\mu/T$ is constant, we also have $\mu u^\mu\nabla_\mu u^\nu + (g^{\nu\mu}{+}u^\nu u^\mu)\nabla_\mu \mu = 0$. Combining these two relations reproduces the first equation of \eqref{EoMs}. Another consequence of the Killing condition is $\nabla_\mu u^\mu = 0$. Together with $\mathfrak{L}_{\beta_\star} n = \mathfrak{L}_{\beta_\star} s = 0$, this yields the conservation laws $\nabla_\mu(nu^\mu)=0$ and $\nabla_\mu(su^\mu)=0$.

\newpage
Evaluating \eqref{theBigMess} at $\epsilon=0$, using \eqref{equilibriamo}, and invoking the equation of motion $F {-}\sigma \nabla_\mu \psi \nabla^\mu \psi =-\nabla_\mu \nabla^\mu \sigma$
, we obtain\footnote{Note that $\phi^\mu(0)\equiv P\beta_\star^\mu$, which tells us that  $\Phi(0)
= -\,T_\infty^{-1}\,\Omega$,
where \(\Omega\) denotes the grandcanonical potential, and \(T_\infty\) is the redshifted temperature measured by an observer at infinity. Upon performing a Wick rotation from thermal to real time \cite[Eq.~(3.17)]{GibbonsHawkingActions1997}, the quantity \(\Phi(0)\) reduces to the action originally employed in \cite{Mitra:2020hbj,Buza:2024jxe} to construct the theory.}
\vspace{-0.1cm}
\begin{equation}\label{theBigMess2}
\begin{split}
\phi^{\mu}(0)={}& \left[p-\tfrac{1}{2}\nabla_\nu \sigma\nabla^\nu \sigma-\tfrac{1}{2} \sigma^2 \nabla_\nu \psi \nabla^\nu \psi \right]\beta_\star^\mu  \, , \\
\Dot{\phi}^{\mu}(0)={}& \beta_\star^\nu \nabla_\nu \Dot{\sigma} \nabla^\mu \sigma-\beta_\star^\mu\nabla_\nu (\Dot{\sigma} \nabla^\nu \sigma)   +\beta_\star^\nu \sigma^2 \nabla_\nu \Dot{\psi} \nabla^\mu \psi-\beta_\star^\mu\sigma^2 \nabla_\nu \Dot{\psi} \nabla^\nu \psi,\\
\Ddot{\phi}^{\mu}(0)={}& \beta_\star^\nu \nabla_\nu \Ddot{\sigma} \nabla^\mu \sigma-\beta_\star^\mu\nabla_\nu (\Ddot{\sigma}\nabla^\nu \sigma)+\beta_\star^\nu \sigma^2 \nabla_\nu \Ddot{\psi} \nabla^\mu \psi-\beta_\star^\mu\sigma^2 \nabla_\nu \Ddot{\psi} \nabla^\nu \psi  \\ 
-{}&\left[\nabla_\nu \Dot{\sigma} \nabla^\nu \Dot{\sigma}+ \Dot{\sigma}^2 \nabla_\nu \psi \nabla^\nu \psi+ 4\sigma \Dot{\sigma} \nabla_\nu \psi \nabla^\nu \Dot{\psi}+\sigma^2 \nabla_\nu \Dot{\psi} \nabla^\nu \Dot{\psi}-\Dot{F}\Dot{\sigma} +\Dot{T}\Dot{s}+\Dot{\mu}\Dot{n}+(Ts{+}\mu n)\Dot{u}_\nu \Dot{u}^\nu \right]\beta_\star^\mu \\
+{}& 2\beta_\star^\nu \nabla_\nu \Dot{\sigma} \nabla^\mu \Dot{\sigma}+2\beta_\star^\nu \nabla_\nu \Dot{\psi}\,\left[2\sigma\Dot{\sigma} \nabla^\mu \psi{+}\sigma^2 \nabla^\mu \Dot{\psi}\right]-2\left[ s\Dot{T}+n\Dot{\mu} \right]\dfrac{\Dot{u}^\mu}{T} .\\
\end{split}
\end{equation}
On the other hand, it is easy to show that
\begin{equation}
\begin{split}
\beta_\star^\nu \nabla_\nu \Dot{\sigma} \nabla^\mu \sigma-\beta_\star^\mu\nabla_\nu (\Dot{\sigma} \nabla^\nu \sigma) ={}& \nabla_\nu(2\Dot{\sigma} \beta_\star^{[\nu} \nabla^{\mu]} \sigma)-\Dot{\sigma}(\mathfrak{L}_{\beta_\star} \nabla\sigma)^\mu \, , \\  
\beta_\star^\nu \nabla_\nu \Ddot{\sigma} \nabla^\mu \sigma-\beta_\star^\mu\nabla_\nu (\Ddot{\sigma} \nabla^\nu \sigma) = {}&\nabla_\nu(2 \Ddot{\sigma} \beta_\star^{[\nu} \nabla^{\mu]} \sigma)-\Ddot{\sigma}(\mathfrak{L}_{\beta_\star} \nabla\sigma)^\mu \, , \\  
\beta_\star^\nu \sigma^2 \nabla_\nu \Dot{\psi} \nabla^\mu \psi-\beta_\star^\mu\sigma^2 \nabla_\nu \Dot{\psi} \nabla^\nu \psi={}& \nabla_\nu(2 \sigma^2 \Dot{\psi}\beta_\star^{[\nu}  \nabla^{\mu]} \psi)-\Dot{\psi}\sigma^2(\mathfrak{L}_{\beta_\star} \nabla\psi)^\mu+\Dot{\psi}\beta_\star^\mu\nabla_\nu(\sigma^2 \nabla^\nu \psi) \, , \\
\beta_\star^\nu \sigma^2 \nabla_\nu \Ddot{\psi} \nabla^\mu \psi-\beta_\star^\mu\sigma^2 \nabla_\nu \Ddot{\psi} \nabla^\nu \psi={}& \nabla_\nu(2\sigma^2 \Ddot{\psi}\beta_\star^{[\nu}  \nabla^{\mu]} \psi)-\Ddot{\psi}\sigma^2(\mathfrak{L}_{\beta_\star} \nabla\psi)^\mu+\Ddot{\psi}\beta_\star^\mu\nabla_\nu(\sigma^2 \nabla^\nu \psi) \, , \\
\end{split}
\end{equation}
where we recall that $\mathfrak{L}_{\beta_\star} \nabla\sigma=\mathfrak{L}_{\beta_\star} \nabla\psi=0$ due to \eqref{equilibriamo}, and that $\nabla_\nu(\sigma^2 \nabla^\nu \psi)=0$, since the state fulfills the equations of motion \eqref{EoMs}. Therefore, we find that $\dot{\phi}^\mu(0)$ can be expressed as the total divergence of the antisymmetric tensor $\mathcal{Z}^{[\mu \nu]}=2\beta_\star^{[\nu} \Dot{\sigma} \nabla^{\mu]} \sigma+2\beta_\star^{[\nu} \sigma^2 \Dot{\psi} \nabla^{\mu]} \psi$. Invoking Stokes' theorem \cite[\S 3.3.3]{PoissonToolkit2009pwt}, and assuming that $\Dot{\sigma}$ and $\Dot{\psi}$ are compactly supported, we obtain
\vspace{-0.1cm}
\begin{equation}\label{gzgzg}
\Dot{\Phi}(0)=\int_{\text{``Cauchy surface''}} \nabla_\nu \mathcal{Z}^{[\mu \nu]} d^3 \mathbf{\Sigma}_\mu =\dfrac{1}{2} \oint_{\text{``Boundary at }\infty\text{''}}  \mathcal{Z}^{[\mu \nu]} d^2 \mathbf{S}_{\mu\nu}=0 \, , 
\end{equation}
which shows that all solutions of \eqref{EoMs} fulfilling \eqref{equilibriamo} are indeed stationary points of $\Phi$. With an analogous argument, one finds that also the first line in the expression \eqref{theBigMess2} for $\Ddot{\phi}^\mu(0)$ can be integrated out, giving $-\frac{1}{2}\Ddot{\Phi}(0)=\int E^\mu d^3 \mathbf{\Sigma}_\mu$, with the following ``information current'' \cite{GavassinoCausality2021}: 
\vspace{-0.1cm}
\begin{equation}\label{infona}
\begin{split}
TE^\mu={}&\big[\nabla_\nu \Dot{\sigma} \nabla^\nu \Dot{\sigma}+ \Dot{\sigma}^2 \nabla_\nu \psi \nabla^\nu \psi+ 4\sigma \Dot{\sigma} \nabla_\nu \psi \nabla^\nu \Dot{\psi}+\sigma^2 \nabla_\nu \Dot{\psi} \nabla^\nu \Dot{\psi}-\Dot{F}\Dot{\sigma}+\Dot{T}\Dot{s}+\Dot{\mu}\Dot{n}+(Ts{+}\mu n)\Dot{u}_\nu \Dot{u}^\nu \big]\dfrac{u^\mu}{2} \\
-{}& u^\nu \nabla_\nu \Dot{\sigma} \nabla^\mu \Dot{\sigma}-u^\nu \nabla_\nu \Dot{\psi}\, \big(2\sigma\Dot{\sigma} \nabla^\mu \psi{+}\sigma^2 \nabla^\mu \Dot{\psi}\big)+\big( s\Dot{T}+n\Dot{\mu}\big)\Dot{u}^\mu .\\
\end{split}
\end{equation}
As a cross-check, we note that the linearized equations of motion $d$\cref{EoMs}$/d\epsilon$ and the equilibrium conditions \cref{equilibriamo} cause the information current to be conserved on-shell, $\nabla_\mu E^\mu=0$, in agreement with the reversible nature of the theory.

\vspace{-0.3cm}
\subsection{Maximality conditions}
\vspace{-0.3cm}

If a configuration satisfying \eqref{equilibriamo} is to represent a genuine local maximum of $\Phi$ (rather than a minimum or a saddle), one must require that $\int E^\mu d^3\mathbf{\Sigma}_\mu = -\tfrac{1}{2}\Ddot{\Phi}(0)\geq 0$ on arbitrary Cauchy surfaces and for all admissible perturbations. This condition is, in essence, equivalent to demanding that the information current $E^\mu$ be future-directed non-spacelike throughout the superfluid. In section \ref{essentially}, we will clarify in what sense this equivalence is only ``approximate''. For now, we simply observe that the requirement that $E^\mu$ be future-directed and non-spacelike is a \textit{sufficient} condition for $-\tfrac{1}{2}\Ddot{\Phi}(0)$ to constitute a non-negative definite functional.

Fix an event $\mathscr{P}$. At that location, $TE^\mu$ may be viewed as a quadratic form in the linearly independent perturbation variables $\{\Dot{T},\Dot{\mu},\Dot{\sigma},\Dot{u}^\nu,\nabla^\nu\Dot{\sigma},\nabla^\nu \Dot{\psi}\}$. We immediately observe that the variables $\nabla^\nu \Dot{\sigma}$ decouple from the rest, and their associated information current is
\begin{equation}\label{thesavior}
TE^\mu_{\{\nabla\Dot{\sigma}\}} =\nabla_\nu \Dot{\sigma}\nabla^\nu \Dot{\sigma} \dfrac{u^\mu}{2} -u^\nu \nabla_\nu \Dot{\sigma}\nabla^\mu \Dot{\sigma}\, ,
\end{equation}
which is always future-directed non-spacelike. Hence, going forward, we can just set $\nabla^\nu \Dot{\sigma}=0$. We also note that, since $dp=sdT+nd\mu+Fd\sigma$, then
\begin{equation}
\begin{split}
\Dot{s}={}& \partial_T s \, \Dot{T}+\partial_T n \, \Dot{\mu}+\partial_T F \, \Dot{\sigma} \, , \\
\Dot{n}={}& \partial_T n \, \Dot{T}+\partial_\mu n\,  \Dot{\mu}+\partial_\mu F\,  \Dot{\sigma} \, , \\
\Dot{F}={}& \partial_T F\,  \Dot{T}+\partial_\mu F\,  \Dot{\mu}+\partial_\sigma F\,  \Dot{\sigma} \, . \\
\end{split}
\end{equation}
Plugging these relations into \eqref{infona}, the terms proportional to $\Dot{T}\Dot{\sigma}$ and $\Dot{\mu}\Dot{\sigma}$ cancel out, leaving
\begin{equation}\label{infona2}
\begin{split}
TE^\mu={}&\big[ \Dot{\sigma}^2 (\nabla_\nu \psi \nabla^\nu \psi {-}\partial_\sigma F)+ 4\sigma \Dot{\sigma} \nabla_\nu \psi \nabla^\nu \Dot{\psi}+\sigma^2 \nabla_\nu \Dot{\psi} \nabla^\nu \Dot{\psi}+\partial_T s \, \Dot{T}^2+2\partial_T n \, \Dot{T}\Dot{\mu}+\partial_\mu n\,  \Dot{\mu}^2+(Ts{+}\mu n)\Dot{u}_\nu \Dot{u}^\nu \big]\dfrac{u^\mu}{2} \\
-{}& u^\nu \nabla_\nu \Dot{\psi}\,\big(2\sigma\Dot{\sigma} \nabla^\mu \psi{+}\sigma^2 \nabla^\mu \Dot{\psi}\big)+\big( s\Dot{T}+n\Dot{\mu}\big)\Dot{u}^\mu .\\
\end{split}
\end{equation}
Working in a local Lorentz frame such that $u^\mu\,{=}\,(1,0,0,0)$, $\nabla^\mu \psi\,{=}\,(\mu,\mu v,0,0)$, and $\partial_3\Dot{\psi}=0$ at $\mathscr{P}$, equation \eqref{infona2} reduces to (setting $\Dot{z}_\mu \equiv\sigma\nabla_\mu \Dot{\psi}$)
\begin{equation}\label{infona3}
\begin{split}
2TE^0={}& \Dot{z}_0^2-[\partial_\sigma F+\mu^2(1{-}v^2)]\Dot{\sigma}^2 +4\mu v\Dot{\sigma} \Dot{z}_1+\Dot{z}_1^2+\Dot{z}_2^2+\partial_T s \, \Dot{T}^2+2\partial_T n \, \Dot{T}\Dot{\mu}+\partial_\mu n\,  \Dot{\mu}^2+(Ts{+}\mu n)\Dot{u}_j \Dot{u}^j\, ,  \\
2TE^1={}&- 2\Dot{z}_0\big(2\mu v \Dot{\sigma}{+}\Dot{z}_1\big)+2\big( s\Dot{T}+n\Dot{\mu}\big)\Dot{u}^1 \, ,\\
2TE^2={}&- 2\Dot{z}_0\,\Dot{z}_2+2\big( s\Dot{T}+n\Dot{\mu}\big)\Dot{u}_2 \, ,\\
2TE^3={}& 2\big( s\Dot{T}+n\Dot{\mu}\big)\Dot{u}_3 \, .\\
\end{split}
\end{equation}
In these coordinates, the requirement that $E^\mu$ be future-directed non-spacelike is equivalent to the requirement that the quadratic form $2T(E^0-\lambda_j E^j)$ be positive definite for all choices of $\lambda_j$ such that $\lambda_j \lambda^j< 1$. The resulting inequalities guarantee that the states \eqref{equilibriamo} are actual maxima of $\Phi$, and thus fulfill conditions \textbf{(b,c)}\footnote{Recall that \eqref{equilibriamo} fulfill condition \textbf{(a)} by construction.}.

\subsubsection{Stability to fluctuations of the order parameter}

One readily verifies that, in every component of $E^\mu$, the ``superfluid'' fluctuations $\{\Dot{\sigma},\Dot{z}_0,\Dot{z}_1,\Dot{z}_2\}$ decouple from the ``normal'' fluctuations $\{\Dot{T},\Dot{\mu},\Dot{u}^j\}$. Consequently, the positivity conditions may be imposed independently on each sector. Focusing first on the superfluid sector, we have the quadratic form
\begin{equation}\label{infosupona}
2T(E^0{-}\lambda_j E^j)_{\text{``superfluid''}}=
\begin{bmatrix}
\Dot{\sigma} & \Dot{z}_0 & \Dot{z}_1& \Dot{z}_2 \\
\end{bmatrix}
\begin{bmatrix}
-\partial_\sigma F-\mu^2(1{-}v^2) & 2\lambda_1\mu v & 2\mu v & 0 \\
2\lambda_1 \mu v & 1 & \lambda_1 & \lambda_2 \\
2\mu v & \lambda_1 & 1 & 0 \\
0 & \lambda_2 & 0 & 1 \\
\end{bmatrix}
\begin{bmatrix}
\Dot{\sigma}\\
\Dot{z}_0 \\
\Dot{z}_1 \\
\Dot{z}_2 \\
\end{bmatrix} \, .
\end{equation}
If we demand that the relevant $4{\times} 4$ matrix be positive definite, we obtain a single inequality (for arbitrary $\lambda_j$):
\begin{equation}\label{stabilitytocondensate}
\partial_\sigma F + \mu^{2}\!\left(1{+}3v^{2}\right) < 0\, .
\end{equation}
If this condition is satisfied everywhere within the superfluid, any configuration obeying \eqref{equilibriamo} constitutes a local maximum of the functional $\Phi$ with respect to variations of the order parameter $\Sigma$. The fact that every choice of $\lambda_j$ leads to the same inequality \eqref{stabilitytocondensate} follows directly from the observation that fluctuations of $\Sigma$ always propagate causally (see Appendix \ref{causalone}), so all observers must agree on whether a state is stable or not \cite{GavassinoSuperluminal2021}\footnote{\label{BigFoot}
Changing $\lambda_j$ amounts to selecting a different normal to the Cauchy surface on which the integral $-\tfrac{1}{2}\Ddot{\Phi}(0)=\int E^\mu d^{3}\mathbf{\Sigma}_\mu$ is evaluated. This is equivalent to changing the reference frame in which conditions \textbf{(b,c)} are examined.}.

\subsubsection{Stability to fluctuations of the normal flow}

Let us now focus on the normal sector, characterized by the fluctuations $\{\Dot{T},\Dot{\mu},\Dot{u}^j\}$. By isotropy, we can set $\Dot{u}^2=\Dot{u}^3=0$, and we are left with the quadratic form
\begin{equation}
2T(E^0{-}\lambda_j E^j)_{\text{``normal''}}=
\begin{bmatrix}
\Dot{T} & \Dot{\mu} & \Dot{u}_1 \\
\end{bmatrix}
\begin{bmatrix}
\partial_T s & \partial_T n & -\lambda_1 s \\
\partial_T n & \partial_\mu n & -\lambda_1 n \\
-\lambda_1 s & -\lambda_1 n & Ts{+}\mu n\\
\end{bmatrix}
\begin{bmatrix}
\Dot{T}\\
\Dot{\mu} \\
\Dot{u}_1 \\
\end{bmatrix} \, .
\end{equation}
Positive definiteness of the relevant $3\times 3$ matrix for all $\lambda_j\in (-1,1)$ produces the following thermodynamic inequalities:
\begin{equation}\label{stabilitytonormal}
\partial_T s >0 \, , \qquad Ts{+}\mu n >0 \, , \qquad \partial_T s \partial_\mu n-(\partial_T n)^2 >0\, ,\qquad  0<c_N^2 \leq 1 \, ,
\end{equation}
where $c_N$ denotes the propagation speed of normal modes, computed in Appendix \ref{causalone}.  
The final inequality is a causality condition, required for \textbf{(b,c)} to remain valid in all reference frames \cite{GavassinoSuperluminal2021} (see footnote \ref{BigFoot}).  
In fact, ensuring that the information current $E^\mu$ is future-directed and non-spacelike is known to enforce linear causality automatically \cite{GavassinoCausality2021}.

If both \eqref{stabilitytocondensate} and \eqref{stabilitytonormal} hold throughout the fluid, the quadratic form $-\tfrac{1}{2}\Ddot{\Phi}(0)$ is positive definite. It follows that the configuration \eqref{equilibriamo} is a local maximum of $\Phi$, satisfies conditions \textbf{(a,b,c)}, and therefore represents an equilibrium state.

\subsection{Landau's critical velocity}\label{LandauIsBack}

We are finally ready to explain the magenta region in figure \ref{fig:Uniform}.

Let us work in Minkowski spacetime and consider a superfluid governed by the equation of state \eqref{EoS}, weakly coupled to a thermal bath characterized by the intensive parameters $\beta_\star^\nu=(1/T_\star,0,0,0)$ and $\alpha_\star=\mu_\star/T_\star$. The spatially homogeneous solutions of \eqref{EoMs} satisfying \eqref{equilibriamo} take the form
\begin{equation}\label{battilemani}
\begin{split}
T &= T_\star \, , \\
\mu &= \mu_\star \, , \\
u^\mu &= (1,0,0,0) \, , \\
\psi &= \text{const} - \mu_\star t + \mu_\star v x \, , \\
\sigma &= \bar{\sigma}(T_\star,\mu_\star,v) \, ,
\end{split}
\end{equation}
where $\bar{\sigma}(T_\star,\mu_\star,v)$ is given in \eqref{nereee}, and we have oriented the $x$--axis along the direction of the superflow. These configurations maximize the pressure with respect to variations of $\sigma$, since $-P(\sigma)$ has a Mexican-hat structure and $\bar{\sigma}$ sits at its minimum. 
For such configurations to represent genuine equilibria, however, one must also ensure stability under \textit{coupled} fluctuations of $\sigma$ and $\psi$. This is immediately guaranteed if the matrix appearing in \eqref{infosupona} is positive definite. This requires that the inequality \eqref{stabilitytocondensate} be satisfied, which, for the equation of state \eqref{EoS}, reduces to
\begin{equation}\label{ohohohoh}
3\lambda \sigma^2 > m_0 (T_c {-} T) + 3 \mu^2 v^2 \, .
\end{equation}
Evaluating this condition on the configuration \eqref{battilemani} yields $v < v_c/\sqrt{3}$, with $v_c = \sqrt{m_0(T_c{-}T)/\mu^2}$. When this bound is violated, the state may become dynamically unstable, and superfluidity may be lost, as suggested in figure \ref{fig:Uniform}.

To prove that the configurations with $v>v_c/\sqrt{3}$ are truly unstable, we need to show that there are indeed fluctuations of $\Sigma$ that increase $\Phi$, so that \eqref{battilemani} is not a local maximum. To this end, let us restrict our attention to the family of solutions of \eqref{EoMs} satisfying, at $t\,{=}\,0$, the conditions $\beta^\nu\,{=}\,\beta_\star^\nu$, $\mu\,{=}\,\mu_\star\,{=}\,\partial^t\psi$, and $\partial_t\sigma\,{=}\,0$. For any such state, the first line of \eqref{theBigMess} implies that $\phi^0=P/T_\star$ at $t=0$. Consequently, the thermodynamic functional $\Phi$ takes the form\footnote{We observe that $\Phi=S+\alpha_\star^IQ_I$ is conserved along any solution of \eqref{EoMs}.  
Hence, it suffices to evaluate $\Phi$ on the initial Cauchy slice at $t=0$. The resulting value is preserved throughout the subsequent evolution.}
\begin{equation}
\Phi=\dfrac{1}{T_\star}\int_{\text{``Slice at }t=0\text{''}}\!\left[
p(T_\star,\mu_\star,0)
+\dfrac{m_0(T_c{-}T_\star)}{2}\sigma^2
-\dfrac{\lambda}{4}\sigma^4
-\dfrac{1}{2}(\partial_1\sigma)^2
-\dfrac{1}{2}\sigma^2(\partial_1\psi)^2
\right]d^3 x\,,
\end{equation}
where we have assumed that any nontrivial gradients point along the $x$--axis. We now expand this class of configurations around the uniform state \eqref{battilemani} by writing 
$\sigma=\bar{\sigma}+\delta\sigma$ and $\partial_1\psi=\mu_\star  v+\partial_1\delta\psi$, 
with $\delta\sigma(x)$ and $\delta\psi(x)$ small and of compact support in $x$. 
To quadratic order in perturbations, the corresponding variation of the thermodynamic functional is
\begin{equation}
\Delta\Phi=-\dfrac{1}{2T_\star}
\int_{\text{``Slice at }t=0\text{''}}\!\left[
2\lambda\bar{\sigma}^2(\delta\sigma)^2
+(\partial_1\delta\sigma)^2
+\bar{\sigma}^2(\partial_1\delta\psi)^2
+4\mu_\star v\,\bar{\sigma}\,\delta\sigma\,\partial_1\delta\psi
\right]d^3 x
+\mathcal{O}(\delta\Sigma^3)\,.
\end{equation}
Consider now the very special class of perturbations satisfying 
$\bar{\sigma}\,\partial_1\delta\psi=-2\mu_\star v\,\delta\sigma$.  
For these configurations, the expression simplifies to
\begin{equation}
\Delta\Phi=-\dfrac{1}{2T_\star}
\int_{\text{``Slice at }t=0\text{''}}\!\left[
2(\lambda\bar{\sigma}^2{-}2\mu_\star^2 v^2)(\delta\sigma)^2
+(\partial_1\delta\sigma)^2
\right]d^3x
+\mathcal{O}(\delta\Sigma^3)\,.
\end{equation}
In the long-wavelength limit, the gradient term $(\partial_1\delta\sigma)^2$ is negligible.  
Since $\lambda\bar{\sigma}^2=\mu_\star^2 (v_c^2{-}v^2)$, the quadratic coefficient scales as 
$\lambda\bar{\sigma}^2-2\mu_\star^2 v^2\propto v_c^2-3v^2$.  
Thus, for $v>v_c/\sqrt{3}$, these perturbations \textit{increase} $\Phi$, and are therefore thermodynamically preferred.
When thermal fluctuations are allowed, such excitations will proliferate and condense, ultimately destroying the superfluid phase.  
Accordingly, $v_c/\sqrt{3}$ plays the role of Landau's critical velocity in this model \cite[\S 23]{landau9}.

We note that, in direct analogy with Landau’s critical-velocity argument, the excitations responsible for the breakdown of the superfluid phase \textit{decrease} the total momentum of the flow (as measured in the rest frame of the bath).  
In fact, for the class of configurations considered here, one has $T^{01}=\mu_\star \sigma^2 \partial_1\psi$ at $t=0$.  
Accordingly, the momentum change induced by a compactly supported perturbation satisfying 
$\bar{\sigma}\,\partial_1\delta\psi=-2\mu_\star v\,\delta\sigma$ is
\begin{equation}
\Delta p^1
=-3\!\int_{\text{``Slice at }t=0\text{''}}
\mu_\star^2\,\delta\sigma^2\,v\, d^3x
+\mathcal{O}(\delta\Sigma^3)\,,
\end{equation}
which is negative, whereas the unperturbed flow is directed along the positive $x$--direction.

\subsection{Behavior of the condensate near hard walls}

Our discussion thus far has tacitly assumed that the superfluid occupies either an unbounded domain or a finite region determined by gravity or by nontrivial spacetime topologies. It is therefore natural to ask how the analysis changes when the fluid is instead confined within a container with perfectly reflecting walls. Standard quantum-mechanical arguments require the condensate amplitude $\sigma$ to always vanish at such boundaries. In what follows, we examine how these boundary conditions modify the maximization of $\Phi$.

To begin with, we note that, for the third condition of \eqref{equilibriamo} to remain valid, the walls themselves must be stationary with respect to the Killing flow generated by $\beta_\star^\nu$. This requirement is physically sensible: if the walls were moving, they would perform work on the ``superfluid + bath'' system, invalidating the argument that leads to the monotonic growth of $\Phi$. Equivalently, thermodynamic equilibrium cannot be attained if the container is being shaken.

Another modification is that the integral in \eqref{PhiasaFlux} no longer extends over all of space, but is restricted to the interior of the container. Consequently, the boundary term in \eqref{gzgzg} is not evaluated at spatial infinity. Instead, we obtain
\begin{equation}
\Dot{\Phi}(0)
= \oint_{\text{``Container walls''}}
\left[\Dot{\sigma} \nabla^{\mu} \sigma
+ \sigma^{2}\Dot{\psi} \nabla^{\mu} \psi \right]\beta_\star^{\nu}
d^{2}\mathbf{S}_{\mu\nu}
 \, .
\end{equation}
Imposing the vanishing of the condensate amplitude at the walls enforces $\sigma=\Dot{\sigma}=0$ in the above integral, ensuring that $\Dot{\Phi}(0)$ continues to vanish. The same conclusion applies to the boundary contributions to $\Ddot{\Phi}(0)$, which also vanish. Thus, the derivation leading to the information current \eqref{infona} proceeds unchanged.

\begin{figure}[b!]
    \centering
\includegraphics[width=0.45\linewidth]{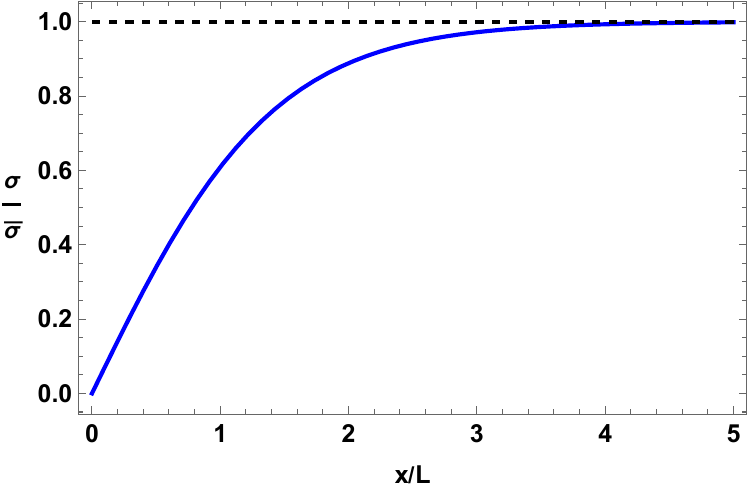}
\caption{Equilibrium profile of the condensate amplitude $\bar{\sigma}(x)$ near a hard wall occupying the region $x\leq 0$, as given by \eqref{tanhaccone}. The condensate vanishes at the boundary and relaxes to its homogeneous value $\bar{\sigma}$ [cf. \eqref{nereee}] over a characteristic distance of order the healing length $L = 1/\sqrt{\lambda \bar{\sigma}^2}$. This expression is formally identical to the standard nonrelativistic result.}
    \label{fig:NearWall}
\end{figure}

Let us now examine how the equilibria \eqref{equilibriamo} behave in a neighborhood of the walls. To this end, we adopt a local Lorentz frame centered at an event $\mathscr{P}$ located on the wall's surface, and such that, locally, $\beta_\star^\nu=(1/T_\star,0,0,0)$, $\text{``Wall''}=\{x\leq 0\}$, and $\nabla^\mu\psi=(\mu_\star,\mu_\star v_x,\mu_\star v_y,0)$. Assuming that all gradients are locally orthogonal to the wall's surface (i.e. point along $x$) and that $\mu_\star$ is approximately constant in this region, the last two lines of \eqref{EoMs} reduce to
\begin{equation}
\begin{split}
&\partial_x(\sigma^2 v_x)=0 \, ,\\
&\partial_x^2\sigma + F + \sigma\mu_\star^2(1 - v_x^2 - v_y^2)=0 \, .
\end{split}
\end{equation}
The first equation implies $\sigma^2 v_x=\text{const}$, and since $\sigma$ vanishes at the wall, it follows that $v_x=0$, i.e.\ the superfluid velocity is tangential to the boundary. Using the equation of state \eqref{EoS}, we are left with
\begin{equation}
\partial_x^2\sigma + \big[m_0(T_c{-}T_\star)-\mu_\star^2 v_y^2\big]\sigma - \lambda\sigma^3 = 0 \, .
\end{equation}
Recalling \eqref{nereee}, namely $\lambda\Bar{\sigma}^2 = m_0(T_c{-}T_\star)-\mu_\star^2 v_y^2$, we introduce the healing length $L = 1/\sqrt{\lambda\Bar{\sigma}^2}$. Defining $f=\sigma/\Bar{\sigma}$ and $\xi=x/L$, the above equation reduces to the dimensionless Gross–Pitaevskii equation $\partial_\xi^2 f + f - f^3 = 0$ \cite[\S 55]{FetterWalecka1971}.
With boundary conditions $f(0)=0$ and $f(+\infty)=1$, the solution yields the familiar condensate profile (see figure \ref{fig:NearWall})
\begin{equation}\label{tanhaccone}
\sigma = \Bar{\sigma}\,\tanh\!\left(\frac{x}{\sqrt{2}\,L}\right) \, .
\end{equation}

\newpage
\subsection{Stability of the boundary layer}\label{essentially}
\vspace{-0.2cm}

A close inspection of Fig.~\ref{fig:NearWall} reveals a thin layer adjacent to the wall in which the inequality~\eqref{ohohohoh} is not satisfied. Consequently, the information current generally ceases to be future-directed non-spacelike in that region. This raises a potential issue: the functional $-\tfrac{1}{2}\Ddot{\Phi}(0)=\int E^\mu d^3\mathbf{\Sigma}_\mu$ might fail to remain positive definite for perturbations of the condensate that are localized near the boundary, thereby opening the possibility of instabilities. The expectation, however, is that the region where $E^0$ becomes problematic is sufficiently narrow that any perturbation strictly confined within it necessarily acquires very large gradients $\nabla_\mu \Dot{\sigma}$; as a result, the gradient contribution~\eqref{thesavior} to the information current overwhelms the negative part, ensuring that $-\tfrac{1}{2}\Ddot{\Phi}(0)$ remains strictly positive. As we demonstrate below, this is indeed the case.

Let us continue to work in the local Lorentz frame introduced in the previous subsection, and assume, for simplicity, that $\Dot{T}=\Dot{\mu}=\Dot{u}^j=\Dot{\psi}=0$. Under these conditions, the time component of the information current reads
\begin{equation}
2TE^0=(\partial_t \Dot{\sigma})^2+\partial_j \Dot{\sigma}\,\partial^j \Dot{\sigma}-(\lambda\Bar{\sigma}^2-3\lambda\sigma^2)\,\Dot{\sigma}^2,
\end{equation}
with $\sigma$ given by \eqref{tanhaccone}. The last term is negative in the region where $\sigma$ is small (i.e., close to the wall). To assess whether this negative contribution can dominate once integrated over the bulk of the superfluid, we set $\partial_t\Dot{\sigma}=\partial_y\Dot{\sigma}=\partial_z\Dot{\sigma}=0$, since these derivatives only increase the integral, and consider the functional\footnote{Here, we examine stability in the local rest frame of the bath. Since fluctuations of the condensate propagate causally, other reference frames must yield the same stability statement \cite{GavassinoSuperluminal2021}. Indeed, the relevant integral \eqref{integruzzone} is the same if we work on other spacelike hypersurfaces. For example, consider the surface $t=\mathfrak{v} x$, with $\mathfrak{v}<1$. Then, dropping the $\partial_y\Dot{\sigma}$ and $\partial_z\Dot{\sigma}$ terms, the integrand becomes $2T(E^0-\mathfrak{v} E^1)=(\partial_t\Dot{\sigma})^2+(\partial_1\Dot{\sigma})^2+2\mathfrak{v}\partial_1\Dot{\sigma}\partial_t\Dot{\sigma}-(\lambda\Bar{\sigma}^2{-}3\lambda\sigma^2)\Dot{\sigma}^2$. However, along the hypersurface, one has $d\Dot{\sigma}/dx=\partial_1\Dot{\sigma}+\mathfrak{v}\partial_t\Dot{\sigma}$, so the kinetic term in the integral becomes $(\partial_t\Dot{\sigma})^2+(\partial_1\Dot{\sigma})^2+2\mathfrak{v}\partial_1\Dot{\sigma}\partial_t\Dot{\sigma}=(1{-}\mathfrak{v}^2)(\partial_t\Dot{\sigma})^2+(d\Dot{\sigma}/dx)^2$. Since the first term is always non-negative, we can drop it, and we recover \eqref{integruzzone}.}
\begin{equation}\label{integruzzone}
-\dfrac{1}{2}\Ddot{\Phi}(0)\propto \int_0^\infty \left\{\left( \dfrac{d\Dot{\sigma}}{dx}\right)^2 -\dfrac{1}{L^2}\left[1-3\tanh^2\!\left(\frac{x}{\sqrt{2}\,L}\right)\right] \Dot{\sigma}^2 \right\} dx\, .
\end{equation}
Making the change of variable $x=\sqrt{2}L\xi$, and using $\tanh^2\xi+\text{sech}^2\xi=1$, this becomes
\begin{equation}
-\dfrac{1}{2}\Ddot{\Phi}(0)\propto \int_0^\infty \left\{\dfrac{1}{2}\left( \dfrac{d\Dot{\sigma}}{d\xi}\right)^2 +\left[2-3\,\text{sech}^2(\xi)\right]\Dot{\sigma}^2 \right\} d\xi\, .
\end{equation}
The integral coincides with the expectation value of the Hamiltonian of a nonrelativistic particle moving in the potential $V(\xi)=2-3\,\text{sech}^2(\xi)$, with the wavefunction supported in $\xi\ge 0$. This is a shifted Pöschl--Teller potential whose lowest eigenvalue is exactly zero \cite[\S 23, Problem 4]{landau3}. Consequently, $-\tfrac{1}{2}\Ddot{\Phi}(0)$ is non-negative definite. In fact, it is strictly positive: the ground state of the Pöschl--Teller potential has nonvanishing support for $\xi<0$, so any admissible $\Dot{\sigma}$ on $\xi\ge 0$ necessarily involves excited states and thus carries strictly positive energy.

The principal insight from the foregoing example is that violations of the inequality \eqref{ohohohoh} may indeed occur locally. However, such violations do not necessarily give rise to instabilities, provided that the region in which they occur is sufficiently thin (its thickness not exceeding the healing length), which is precisely what we observe near boundaries.

\vspace{-0.2cm}
\section{Some notable equilibrium configurations}
\vspace{-0.2cm}

In the preceding section, we established that any field configuration $\{\beta^\mu(x^\alpha),\mu(x^\alpha),\Sigma(x^\alpha)\}$ satisfying
\begin{equation}\label{EquillLLibria}
\begin{split}
& \beta^\mu = \text{``Killing vector''}\,, \\
& \frac{\mu}{T} = \text{``Constant''}=-\beta^\mu \nabla_\mu \psi\,, \\
& \mathfrak{L}_{\beta}\text{``Observables''}=0\,, \\
& \nabla_\mu (\sigma^2 \nabla^\mu \psi)=0\,, \\
& \nabla_\mu \nabla^\mu \sigma + F - \sigma \nabla_\mu \psi \nabla^\mu \psi = 0
\end{split}
\end{equation}
constitutes a solution of the equations of motion \eqref{EoMs}, is stable under small fluctuations, and serves as a local late-time attractor (provided certain thermodynamic inequalities are satisfied away from boundaries). In what follows, we construct such configurations in a few simple geometrical settings, always assuming the equation of state \eqref{EoS}, and always working in Minkowski spacetime.

\newpage
\vspace{-0.2cm}
\subsection{Vortex solution}\label{vortexco}
\vspace{-0.2cm}

Working in Minkowski spacetime, we adopt cylindrical coordinates $\{t,r,\varphi,z\}$, in which the line element takes the form
$ds^{2} = -dt^{2} + dr^{2} + r^{2} d\varphi^{2} + dz^{2}$.
We seek cylindrically symmetric configurations with $\beta^{\mu}\partial_{\mu}=T_\star^{-1}\partial_{t}$ and $\mu/T=\mu_\star/T_\star$, where $T_\star$ and $\mu_\star$ are constants, ensuring that $\beta^\mu$ is a Killing vector. In this setting, the Josephson relation reduces to $\partial_{t}\psi=-\mu_\star$. Furthermore, axial symmetry together with the periodic identification $\varphi\sim\varphi+2\pi$ implies
$\psi(\varphi)=\psi(0)+\ell\varphi$, with $\ell\in\mathbb{Z}$. Note that $\ell$ cannot depend on $r$, as that would introduce a discontinuity. Additionally, invariance under $z$--translations implies that $\partial^2_z\psi=\partial_z\partial_r\psi=0$, and thus we have
\begin{equation}
\psi=\varpi(r)-\mu_\star t+\varphi \ell +\kappa z\, .
\end{equation}
Here, the constant $\kappa$ represents a uniform superflow along the $z$–direction, whose presence is compatible with the imposed symmetries. Finally, cylindrical symmetry implies that the condensate is a function of $r$ only, $\sigma =\sigma(r)$. Under these assumptions, the equations governing the condensate take the form
\begin{equation}
\begin{split}
& \partial_r (r \sigma^{2} \varpi') = 0 \,, \\
& \frac{1}{r}\partial_r (r \partial_r \sigma)
  + \left[m_0 (T_c{-}T_\star) - \kappa^{2} - \frac{\ell^{2}}{r^{2}} - (\varpi')^{2} \right]\sigma
  - \lambda \sigma^{3} = 0 \, .
\end{split}
\end{equation}
The first equation integrates to $r \sigma^{2} \varpi' = C$ for some constant $C$. Evaluating this relation at $r=0$ forces $C=0$, and thus $\varpi'=0$. Substituting this into the second equation yields
\begin{equation}\label{galactic}
\frac{1}{\xi}\partial_\xi(\xi \partial_\xi f)
  + \left( 1 - \frac{\ell^{2}}{\xi^{2}} \right) f - f^{3} = 0 \, ,
\end{equation}
where we have introduced the dimensionless quantities $f=\sigma/\Bar{\sigma}$ and $\xi=r/L$. Here, $\Bar{\sigma}$ denotes the asymptotic value of the condensate far from $r=0$, determined by
$\lambda \Bar{\sigma}^{2} = m_0 (T_c{-}T_\star) - \kappa^{2}$, and
$L = 1/\sqrt{\lambda\Bar{\sigma}^{2}}$ is the healing length. This is exactly the same equation that governs the structure of a vortex in the Newtonian theory. The resulting profile of $f$ is very similar to that in figure \ref{fig:NearWall} (see \cite[\S 30]{landau9}): the condensate vanishes at $r=0$ and approaches $\Bar{\sigma}$ over a couple of healing lengths.

\subsection{Condensate profile across the section of a cylindrical annulus}

\begin{figure}[h!]
    \centering
\includegraphics[width=0.46\linewidth]{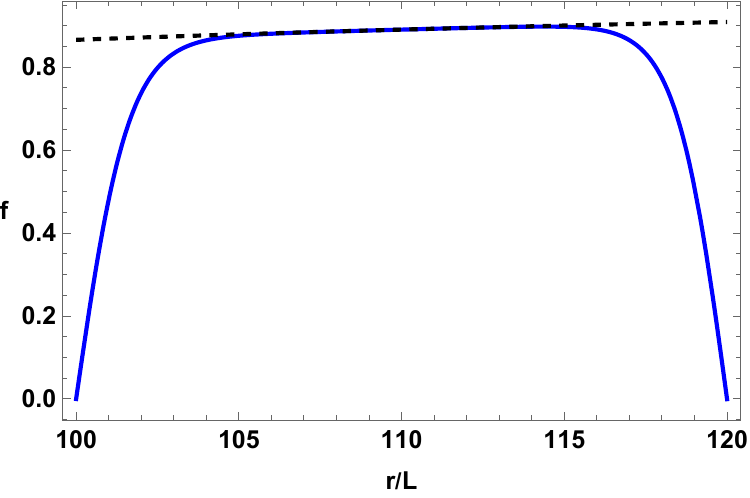}
\caption{Solution $f(\xi)$ of \eqref{galactic} with $\ell=50$, and boundary conditions 
$f(100)=f(120)=0$
. The dashed line is the function $\sqrt{1-\ell^2/\xi^2}$, corresponding to the uniform prediction \eqref{nereee}, which is the value of the condensate magnitude employed in Landau's two-fluid model.}
    \label{fig:Annular}
\end{figure}

Consider a superfluid confined to circulate between two static, concentric cylinders of radii $r_1$ and $r_2$. Treating the cylinders as perfectly reflecting walls, the condensate amplitude must vanish at $r=r_1$ and $r=r_2$, and recover the value \eqref{nereee} (corresponding to the uniform phase) a few healing lengths away from the boundaries. Since the geometry is still cylindrical, equation \eqref{galactic} continues to apply. An illustrative example is shown in figure \ref{fig:Annular}, for a configuration in which the condensate phase carries a winding number $\ell = 50$ around the annulus. In this example, the annular region has a thickness of $20$ healing lengths, while its inner radius is $100$ healing lengths.

\subsection{Superfluids on an accelerating rocket}
\vspace{-0.3cm}

\begin{figure}[h!]
    \centering
\includegraphics[width=0.45\linewidth]{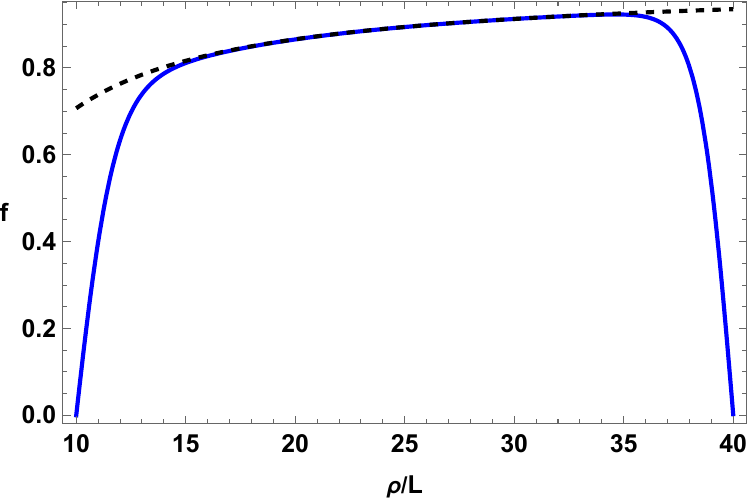}
\includegraphics[width=0.45\linewidth]{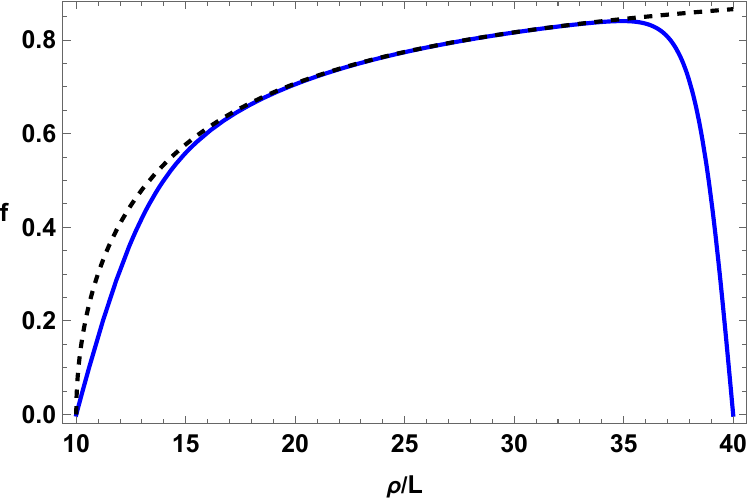}
\includegraphics[width=0.45\linewidth]{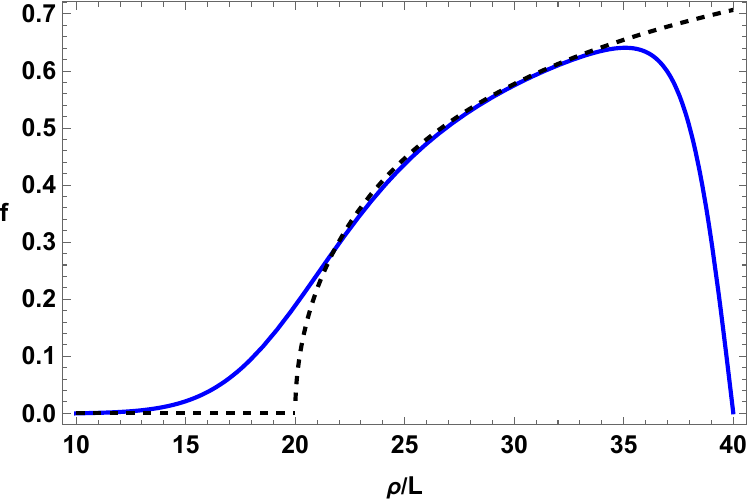}
\includegraphics[width=0.45\linewidth]{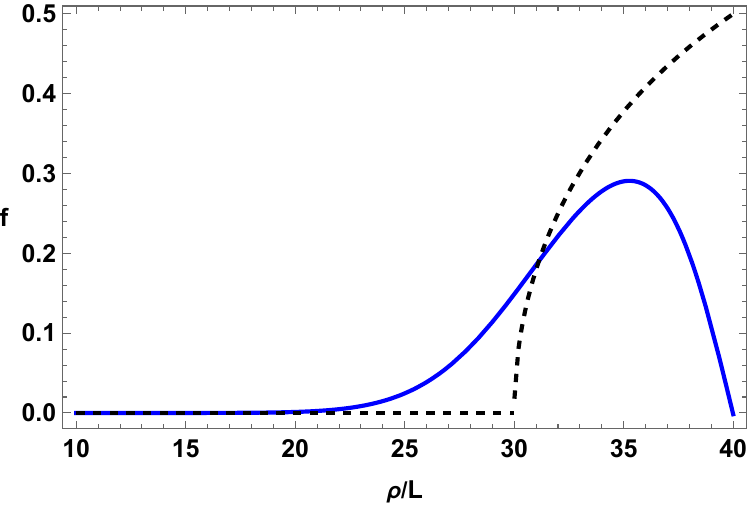}
\caption{Solutions of \eqref{visually} for the condensate profile of a superfluid in equilibrium between two uniformly accelerating walls, located at $\varrho_{1}=10\,L$ and $\varrho_{2}=40\,L$. The four panels correspond to $\xi_\star = 5$ (upper left), $10$ (upper right), $20$ (lower left), and $30$ (lower right). The dashed curve represents $\sqrt{1-\xi_\star/\xi}$, i.e.\ the uniform prediction \eqref{nereee}, which is the condensate magnitude entering Landau's two--fluid model.}
\label{fig:Rocket}
\end{figure}

We now examine a superfluid contained within a bucket undergoing hyperbolic motion. We orient our axis so that the proper acceleration points in the $x$--direction, and introduce Rindler's coordinates $\{\tau,\varrho,y,z\}$, defined through $t=\varrho\sinh(a\tau)$ and $x=\varrho\cosh(a\tau)$. The metric then takes the form $ds^{2}=-a^{2}\varrho^{2}\, d\tau^{2}+d\varrho^{2}+dy^{2}+dz^{2}$, and the front and back walls of the bucket lie on some hyperbolae $\varrho\,{=}\,\varrho_1$ and $\varrho\,{=}\,\varrho_2$, respectively. In this geometry, we may choose $\beta^\mu \partial_\mu\, {=}\, T_\star^{-1}\partial_\tau$ and $\mu/T\,{=}\,\mu_\star/T_\star$,
with $T_\star$ and $\mu_\star$ constant, so that $\beta^\mu$ is a Killing field. Thus, the Josephson relation reads $\partial_\tau\psi\,{=}\,{-}\mu_\star$, and we can set $\psi\,{=}\,\varpi(\varrho)\,{-}\,\mu_\star \tau$. We are thus led to solve the differential equations
\vspace{-0.1cm}
\begin{equation}
\begin{split}
& \partial_\varrho (\varrho \sigma^2 \varpi')=0\,, \\
& \dfrac{1}{\varrho}\partial_\varrho( \varrho\partial_\varrho \sigma) + \left[m_0\left(T_c-\dfrac{T_\star}{a\varrho}\right) -(\varpi')^2\right]\sigma-\lambda\sigma^3 = 0\, .\\
\end{split}
\end{equation}
The first relation integrates to $\varrho\sigma^2\varpi'=C$. Recalling that $\sigma$ vanishes at $\varrho_1$ and $\varrho_2$, we must force $C=0$, and therefore $\varpi'=0$ throughout the domain. Introducing the zero-temperature condensate amplitude and healing length via $\lambda\Bar{\sigma}^2=m_0 T_c$ and $L=1/\sqrt{\lambda\Bar{\sigma}^2}$, the second equation reduces to the dimensionless form
\vspace{-0.1cm}
\begin{equation}\label{visually}
\dfrac{1}{\xi}\partial_\xi( \xi \partial_\xi f) + \left(1-\dfrac{\xi_\star}{\xi}\right) f- f^3 = 0\, ,
\end{equation}
where we have defined $f=\sigma/\Bar{\sigma}$, $\xi=\varrho/L$, and $\xi_\star=T_\star/(aLT_c)$. In figure \ref{fig:Rocket}, we provide some solutions for the same values of $\varrho_1$ and $\varrho_2$, while increasing $\xi_\star$. We see that, as implied by the Tolman law ($T \sim 1/\varrho$), the fluid may remain superfluid near the rocket’s front and become normal near its rear, with a phase transition in between. Whereas Landau’s two-fluid model yields a non-differentiable change from $\sigma=0$ for $T>T_c$ to $\sigma>0$ for $T<T_c$, the Gross–Pitaevskii description instead produces a smooth crossover over a few healing lengths.

\subsection{Vortex-free superfluid in a rotating bucket}
\vspace{-0.2cm}

\begin{figure}[h!]
    \centering
\includegraphics[width=0.46\linewidth]{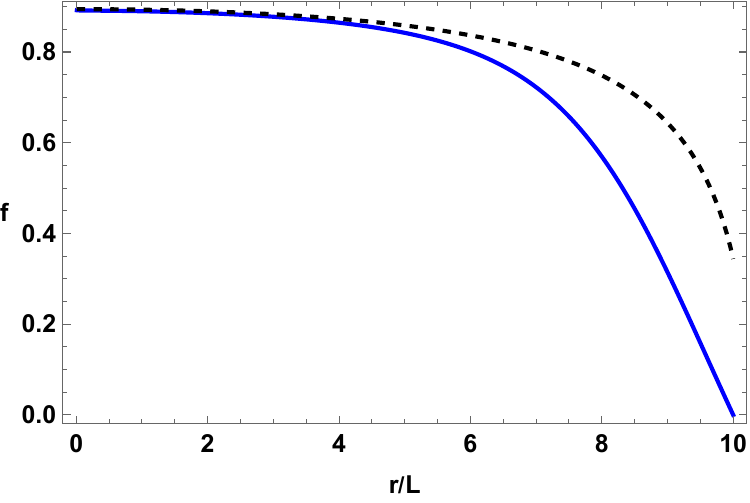}
\includegraphics[width=0.46\linewidth]{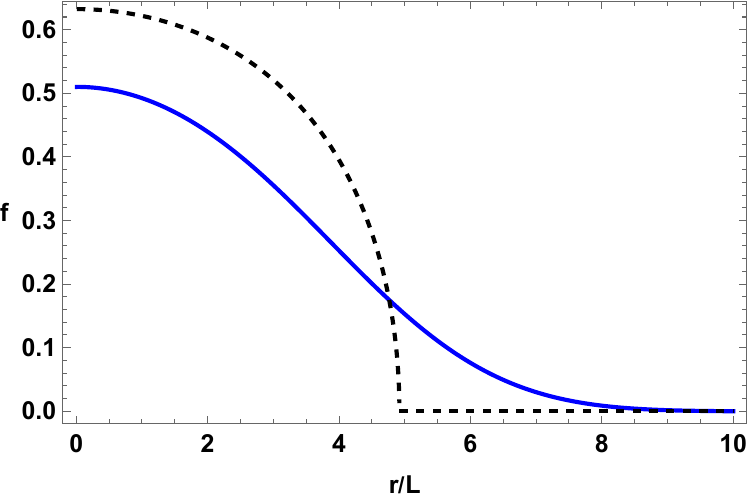}
\includegraphics[width=0.46\linewidth]{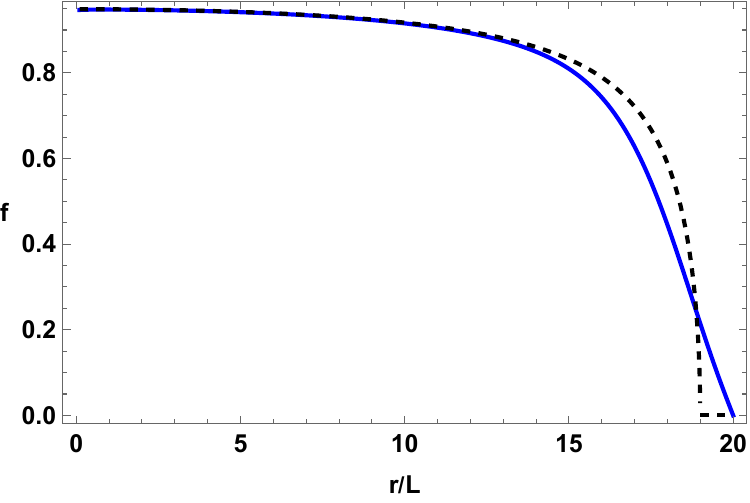}
\includegraphics[width=0.46\linewidth]{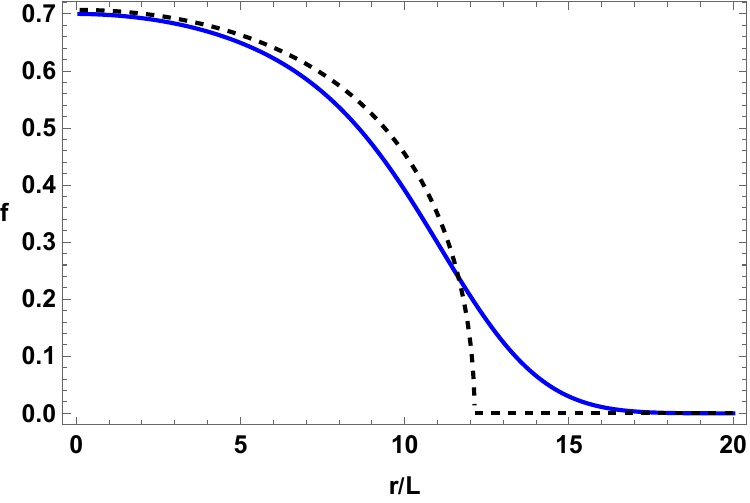}
\caption{Solutions of \eqref{biggone} describing the condensate amplitude for a superfluid in equilibrium within a rotating cylindrical container. The four panels correspond respectively to the parameter sets $\{r_{\text{Bucket}}/L,\theta,\nu,\omega\}=\{10,1/12,0.2,0.1\}$ (upper left), $\{10,1/11,0.6,1\}$ (upper right), $\{20,1/21,0.1,0.1\}$ (lower left), and $\{20,1/21,0.5,0.5\}$ (lower right). The dashed curve represents the uniform prediction \eqref{nereee}, which is the condensate magnitude entering Landau's two--fluid model.}
    \label{fig:RotateBucket}
\end{figure}

As a final illustration, we examine a superfluid confined within a cylindrically symmetric container rotating at fixed angular velocity $\Omega$. In equilibrium, the inverse--temperature four--vector of the superfluid must equal the Killing field $\beta^\mu \partial_\mu = T_\star^{-1}(\partial_t + \Omega\,\partial_\varphi)$, with $T_\star=\text{const}$. Fixing $\mu/T=\mu_\star/T_\star$, where $\mu_\star$ is constant, the Josephson relation becomes $(\partial_t + \Omega\,\partial_\varphi)\psi = -\mu_\star$. In the absence of vortices, $\psi$ is independent of $\varphi$ and may thus be written as $\psi = \varpi(r) - \mu_\star t$. The condensate variables then satisfy the differential equations
\begin{equation}
\begin{split}
& \partial_r (r\sigma^2 \varpi')=0\,, \\
& \dfrac{1}{r}\partial_r(r \partial_r \sigma) + \left[m_0 (T_c-T)-\mu^2+\mu_\star^2-(\varpi')^2 \right]\sigma -\lambda\sigma^3 = 0\\
\end{split}
\end{equation}
Again, the first equation gives $\varpi'=0$. Hence, defined $\lambda\Bar{\sigma}^2=m_0 T_c$ and $L=1/\sqrt{\lambda\Bar{\sigma}^2}$, the second equation reduces to
\begin{equation}\label{biggone}
\dfrac{1}{\xi}\partial_\xi(\xi \partial_\xi f) +\left(1-\dfrac{\theta}{\sqrt{1-\omega^2 \xi^2}}-\dfrac{\nu\,\omega^2 \xi^2}{1-\omega^2 \xi^2} \right) f - f^3 = 0\, ,
\end{equation}
where we have introduced the dimensionless quantities $f\,{=}\,\sigma/\Bar{\sigma}$, $\xi\,{=}\,r/L$, $\omega\,{=}\,\Omega L$, $\theta\,{=}\,T_\star/T_c$, and $\nu\,{=}\,\mu_\star^2/(m_0 T_c)$. The above equation must be solved with boundary conditions $f'(0)=0$ (smoothness on the axis) and $f(\text{``Bucket wall''})=0$. Representative solutions for different parameter choices are shown in Fig.~\ref{fig:RotateBucket}. The qualitative behavior mirrors that of the preceding example: the Tolman law $T\sim 1/\sqrt{1{-}\Omega^{2}r^{2}}$ may keep the central region in the superfluid phase while driving the outer layers normal, thereby generating a phase transition in between. As before, Landau’s two-fluid model predicts a non-differentiable change from $\sigma=0$ for $T>T_c$ to $\sigma>0$ for $T<T_c$, whereas the Gross--Pitaevskii framework yields a smooth crossover occurring over several healing lengths.

\newpage
\section{Conclusions}
\vspace{0.2cm}

In this work, we have employed the relativistic (finite-temperature) Gross--Pitaevskii--type formulation of \cite{Mitra:2020hbj,Buza:2024jxe} to construct and analyze stationary configurations of relativistic superfluids in regimes where the condensate amplitude exhibits appreciable spatial variation. This framework promotes the condensate magnitude to an independent degree of freedom, enabling a controlled treatment of boundary layers, vortex cores, and configurations in which thermal or kinematic gradients drive the system toward the superfluid--normal phase boundary.

First, we identified the general conditions under which a stationary configuration represents a local maximum of the thermodynamic functional $\Phi=S+\alpha_\star^I Q_I$, and is therefore Lyapunov stable and dynamically admissible. This analysis led to a fully relativistic derivation of Landau's critical velocity: for the Mexican--hat potential adopted here, superfluidity breaks down at $v=v_c/\sqrt{3}$, where $v_c$ is the critical velocity inferred from the uniform-phase condensate. Perturbations violating this bound were shown to increase $\Phi$, confirming the onset of an instability in the strict thermodynamic sense.

Second, we demonstrated that several canonical Newtonian results remain unchanged in the relativistic theory. Both the vortex profile and the near-wall boundary layer reproduce the standard Gross--Pitaevskii forms, including their characteristic healing-length scaling. The stability of the boundary layer was analyzed explicitly by mapping the relevant fluctuation functional to a shifted Pöschl--Teller Hamiltonian, whose spectrum ensures positivity of the second variation of $-\Phi$, despite local violations of the stability inequalities dictated by the information-current technique.

Finally, we investigated equilibrium superflows in genuinely relativistic geometries. In accelerating and rotating containers, Tolman temperature gradients (and their rotational analogues) were shown to generate spatially varying effective temperatures that can drive continuous transitions from superfluid to normal phases across the domain. Whereas Landau's two-fluid model predicts a non-differentiable change across the transition, the relativistic Gross--Pitaevskii description yields a smooth crossover occurring over several healing lengths. One can then identify the onset of criticality at the pseudo-critical value, $\xi_{\rm pc},$ by considering when the derivative of the condensate with respect to $\xi$ attains a maximum value. This is in contrast to Landau theory, where the derivative diverges at $\xi_\star.$ For the example of the bottom right panel of \cref{fig:Rocket}, we see that $\xi_{\rm pc}\sim 20.8$, while Landau theory has $\xi_\star=20.$ This situation is analogous to the pseudo-critical temperature in QCD, where the quark mass explicitly breaks chiral symmetry, leading to a shift from the critical temperature of the massless theory to the pseudo-critical temperature \cite{HotQCD:2019xnw}. As a future direction, it would be interesting to extract the universal features of the accelerating superfluid, e.g.~the associated critical exponents.

Taken together, our results confirm that the relativistic Gross--Pitaevskii framework offers a consistent and predictive description of relativistic superfluids near critical surfaces. It reproduces the expected Newtonian limits, provides a sharp analytic criterion for Landau's critical velocity, and remains stable in the presence of geometrically induced boundary layers. More broadly, this analysis illustrates the utility of the framework for studying relativistic superfluid equilibria in general spacetimes, and paves the way for future investigations of neutron-star superfluidity.

\vspace{0.2cm}
\section*{Acknowledgements}

LG is supported by a MERAC Foundation prize grant,  an Isaac Newton Trust Grant, and funding from the Cambridge Centre for Theoretical Cosmology. AS is supported by funding from the Horizon Europe research and
innovation programme under the Marie Sklodowska-Curie grant agreement No. 101103006,
the project N1-0245 of Slovenian Research Agency (ARIS) and financial support through
the VIP project UNLOCK under contract no. SN-ZRD/22-27/510.

\appendix

\vspace{-0.2cm}
\section{Causal cone of the theory}\label{causalone}
\vspace{-0.2cm}

Consider the non-linear system of partial differential equations \eqref{EoMs}.  
We take $\{T,\mu,u^\nu,\psi,\sigma\}$ as the fundamental variables and aim to compute the associated characteristic polynomial $\mathcal{P}(\xi)$, which defines the local cone within which information propagates.  
Using the Leray–index method \cite{DisconziReview:2023rtt}, one finds that $\psi$ and $\sigma$ propagate on the light cone, so that the contribution of the second and third lines of \eqref{EoMs} factorizes as $\mathcal{P}(\xi) = (g^{\mu\nu}\xi_\mu\xi_\nu)^{2}\mathcal{Q}(\xi)$ \cite{Causality_bulk}.  
Clearly, the light cone is a causal characteristic. We can thus focus on the remaining factor $\mathcal{Q}(\xi)$, which governs the propagation of $T$, $\mu$, and $u^\nu$. This reads (defined $\Delta^{\mu \nu}=g^{\mu \nu}{+}u^\mu u^\nu$)
\newpage
\begin{equation}
\begin{split}
\mathcal{Q}(\xi)={}&\det 
\begin{bmatrix}
\partial_T s\, u^\mu\xi_\mu & \partial_\mu s\, u^\mu\xi_\mu & s \Delta^\mu_\nu\xi_\mu \\
\partial_T n\, u^\mu\xi_\mu & \partial_\mu n\, u^\mu\xi_\mu & n \Delta^\mu_\nu\xi_\mu \\
s\Delta^{\alpha\mu}\xi_\mu & n \Delta^{\alpha\mu}\xi_\mu & (Ts{+}\mu n)g^\alpha_\nu u^\mu\xi_\mu \\
\end{bmatrix} = (u^\mu \xi_\mu)^{-2} \det 
\begin{bmatrix}
\partial_T s & \partial_\mu s & s \Delta^\mu_\nu\xi_\mu \\
\partial_T n & \partial_\mu n & n \Delta^\mu_\nu\xi_\mu \\
s\Delta^{\alpha\mu}\xi_\mu & n \Delta^{\alpha\mu}\xi_\mu & (Ts{+}\mu n)g^\alpha_\nu (u^\mu\xi_\mu)^2 \\
\end{bmatrix}\\
={}&(u^\mu \xi_\mu)^{-2} \det 
\begin{bmatrix}
\partial_T s & \partial_\mu s  \\
\partial_T n & \partial_\mu n  \\
\end{bmatrix}
\det\left\{(Ts{+}\mu n)g^\alpha_\nu (u^\mu\xi_\mu)^2- \begin{bmatrix}
s & n \\    
\end{bmatrix}
\begin{bmatrix}
\partial_T s & \partial_\mu s  \\
\partial_T n & \partial_\mu n  \\
\end{bmatrix}^{-1}
\begin{bmatrix}
s\\
n\\
\end{bmatrix}\Delta^{\alpha\mu}\xi_\mu \Delta^\lambda_\nu \xi_\nu\right\} \\
={}& (u^\mu \xi_\mu)^{4}  
(\partial_T s\partial_\mu n{-}\partial_T n \partial_\mu s)(Ts{+}\mu n)^4
\left\{ (u^\mu\xi_\mu)^2- \dfrac{1}{Ts{+}\mu n}\begin{bmatrix}
s & n \\    
\end{bmatrix}
\begin{bmatrix}
\partial_T s & \partial_\mu s  \\
\partial_T n & \partial_\mu n  \\
\end{bmatrix}^{-1}
\begin{bmatrix}
s\\
n\\
\end{bmatrix}\Delta^{\mu\nu}\xi_\mu \xi_\nu\right\} \, . \\
\end{split}
\end{equation}
The characteristics associated with the term $(u^\mu \xi_\mu)^4$ describe pure advection, and they are always causal, since $u^\mu$ is timelike. The characteristic manifold associated with the large graph bracket is an acoustic cone, with characteristic sound speed
\begin{equation}
c_N^2=\dfrac{1}{Ts{+}\mu n}\begin{bmatrix}
s & n \\    
\end{bmatrix}
\begin{bmatrix}
\partial_T s & \partial_\mu s  \\
\partial_T n & \partial_\mu n  \\
\end{bmatrix}^{-1}
\begin{bmatrix}
s\\
n\\
\end{bmatrix} \, .
\end{equation}
This should be interpreted as the speed of propagation of sound waves in the normal component. Causality demands $0<c_N^2\leq 1$.

\bibliography{Biblio}

\label{lastpage}

\end{document}